\documentclass{mem}
\usepackage{natbib}\usepackage{txfonts}\usepackage{balance}
\usepackage{graphicx}
\usepackage[a4paper,breaklinks,dvipdfm]{hyperref}
\idline{75}{282}
\begin{document}
\def\teff{$T\rm_{eff }$}
\def\kms{$\mathrm {km s}^{-1}$}
\def\glog {log\,g}
\def\logg {log\,g}
\newcommand{\xx}{\ensuremath{\mathrm{1D}_{\mathrm{LHD}}}}
\newcommand{\mD}{\ensuremath{\left\langle\mathrm{3D}\right\rangle}}
\newcommand{\mlp}{\ensuremath{\alpha_{\mathrm{MLT}}}}
\newcommand{\LHD}{{\sf LHD}}
\newcommand{\cobold}{{\sf CO$^5$BOLD}}
\newcommand{\linfor}{{\sf Linfor3D}}
\newcommand{\Nt}{\ensuremath{N_\mathrm{t}}}
\newcommand{\tchar}{\ensuremath{t_\mathrm{c}}}

\title{
Effects of granulation on neutral copper  resonance lines in metal-poor stars
}
\author{
P. \,Bonifacio\inst{1,2,3} 
\and
E. Caffau\inst{2}
\and
H.-G. Ludwig  \inst{1,2}
          }
  \offprints{P. Bonifacio}
\institute{
CIFIST Marie Curie Excellence Team
\and
GEPI, Observatoire de Paris, CNRS, Universit\'e Paris Diderot; Place
Jules Janssen 92190
Meudon, France
\and
Istituto Nazionale di Astrofisica --
Osservatorio Astronomico di Trieste, Via Tiepolo 11,
I-34131 Trieste, Italy
}

\authorrunning{Bonifacio, Caffau \& Ludwig }

\titlerunning{Effects of granulation on Cui{\sc I} lines}

\abstract{
We make use of three dimensional hydrodynamical
simulations to investigate the effects of granulation
on the \ion{Cu}{i}  lines of Mult.\,1 \relax  in the near UV, at 
324.7\,nm and 327.3\,nm.
These lines remain strong even at very low metallicity
and provide the opportunity to study the chemical
evolution of Cu in the metal-poor populations.
We find very strong granulation effects on these lines.
In terms of abundances the neglect of such effects can lead to an
overestimate of the A(Cu) by as much as 0.8 dex in dwarf
stars.
Comparison of our computations with stars in the
metal-poor Globular Clusters NGC 6752 and NGC 6397,
show that there is a systematic discrepancy between 
the copper abundances derived from Mult.\,  \relax 2
in TO stars
and those derived in giant stars of the same cluster
from the lines of Mult.\,2 at at 510.5\,nm and 587.2\,nm.
We conclude that the \ion{Cu}{i} resonance lines are not
reliable indicators of Cu abundance and we believe
that an investigations of departures from LTE is mandatory
to make use of these lines.
\keywords{
Hydrodynamics -- Line: formation -- Stars: abundances -- Galaxy: globular clusters --  NGC~6397,
NGC~6752
}
}
\maketitle{}

\begin{figure*}
\resizebox{\hsize}{!}{\includegraphics[clip=true]{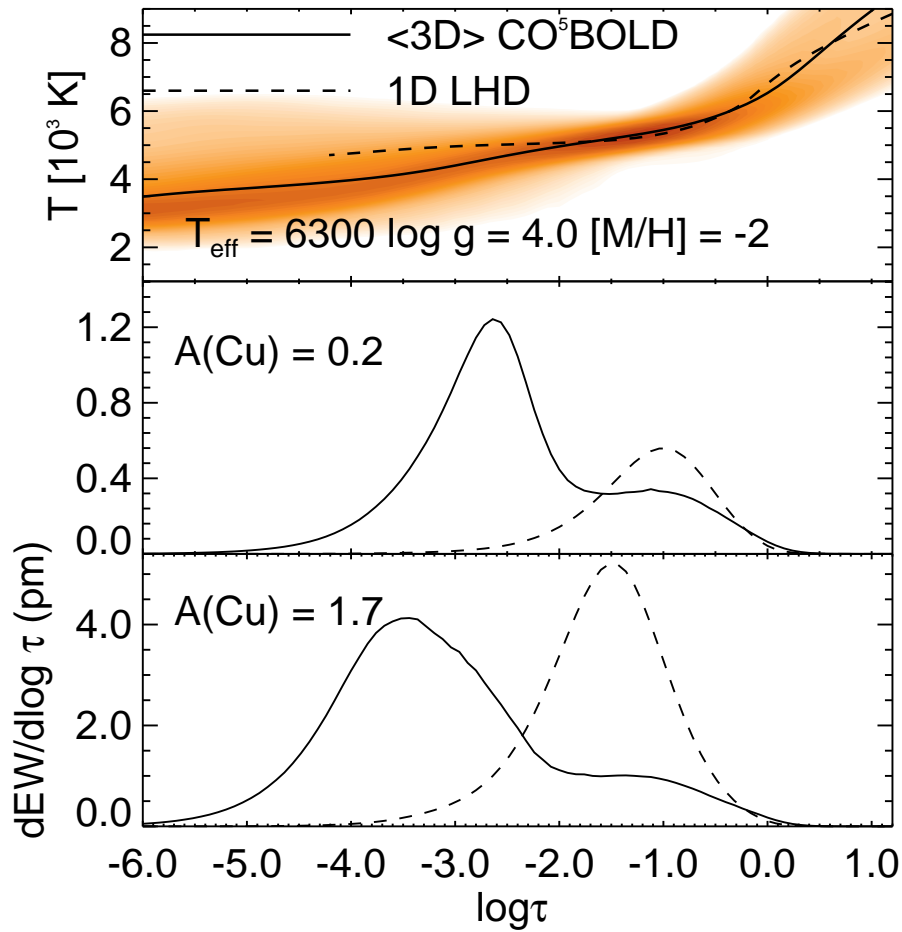}}
\caption{\footnotesize
The top panel shows the temperature distribution
for the hydrodynamical model
d3t63g40mm20n01. The shading allows
to visualise the temperature histogram of the model
(over space and time), for any given optical depth.
A darker colour indicates a larger number of cells in 
the given temperature bin.   
In the two lower panels we show the
contribution functions  of the EW at disc-centre,
defined such that their integral over $\log\tau_\lambda$
gives the EW \citep{magain86}, for the \ion{Cu}{i}
324.7\,nm line and the model d3t63g40mm20n01 for two different
values of Cu abundance. In the upper panel A(Cu)=0.2, in the
lower panel A(Cu)=1.7. The solid lines refer to the 3D model,
the dashed lines to the corresponding \xx\ model. 
}
\label{ttau}
\end{figure*}

\begin{table*}
\caption{Atmospheric parameters and copper abundances for the program stars.\label{abun}}
\begin{center}
\begin{tabular}{lcccccccc}
\hline\hline\noalign{\smallskip}
Star  & T$_{\rm eff}$ & $\log g$   & [Fe/H] & $\xi$ &     A(Cu)  & $\sigma$ & A(Cu)  & $\sigma$\\
      &     K         & [cgs]      &  dex   & \kms &  \multicolumn{2}{c}{1D} &\multicolumn{2}{c}{3D} \\
\hline\noalign{\smallskip}
Cl* NGC 6752 GVS  4428     & 6226      &  4.28  & -1.52 & 0.70 & 3.23 & 0.08 & 2.56 & 0.16 \\
Cl* NGC 6752 GVS 200613    & 6226      &  4.28  & -1.56 & 0.70 & 3.01 & 0.05 & 2.23 & 0.07 \\
Cl* NGC 6397 ALA 1406      & 6345      &  4.10  & -2.05 & 1.32 & 1.33 & 0.03 & 0.74 & 0.05 \\
Cl* NGC 6397 ALA 228       & 6274      &  4.10  & -2.05 & 1.32 & 1.30 & 0.03 & 0.73 & 0.05\\
Cl* NGC 6397 ALA 2111      & 6207      &  4.10  & -2.01 & 1.32 & 1.19 & 0.02 & 0.60 & 0.02 \\
HD 218502          & 6296      &  4.13  & -1.85 & 1.00  & 1.52 & 0.09 & 0.95 & 0.04 \\
\hline
\end{tabular}
\end{center}

\end{table*}

\section{Introduction}

In our quest to understand how the Universe
evolved from the primordial chemical
composition, consisting of H, He, and traces of Li,
to the present day complexity, we strive to uncover
all the nucleosynthetic channels. This requires to 
measure the evolution of as many chemical species 
as possible. For some species this becomes difficult
at low metallicities, when all the observable lines become
very weak. This is the case for Cu, the measurements
of Cu abundances are mainly based on the 
lines of Mult. 2 at 510.5\,nm and 578.2\,nm   
\citep{Sneden91,Mishenina}. However, such lines
become very weak at low metallicities, even in cool giants.
The strongest line, at  510.5\,nm, has an equivalent
with of a few tenths of picometer for a K giant of
metallicity --2.5 and becomes very difficult to measure.
This induced several groups \citep{Bihain,Cohen08}
to push the observations in the near UV, where the
 \ion{Cu}{i} resonance lines
at 324.7\,nm and 327.3\,nm are stronger by  a factor of ten
and can be measured at the lowest metallicities.
These lines are formed in the cool outer layers of
the stellar atmosphere.
Such layers are formally stable
against convection and
1D model atmospheres cannot account for
the effects of convective motions (``granulation'') here.
The use of 3D hydrodynamical simulations
has shown that one effect of this is a
steeper temperature gradient in the outer layers
than what predicted by 1D models \citep{A99,A05}.
This effect is often referred to as ``overcooling''
and is more pronounced for metal-poor stars.
In this contribution we wish to investigate
the effects of granulation on the formation
of the \ion{Cu}{i} resonance lines
in metal-poor stars.
To do so we analyse the lines in the turn-off
stars of two metal-poor Globular Clusters
(NGC 6752 and NGC 6397) and compare the
derived abundances, both in 1D and 3D,
with those derived for giants, making use of 
the lines of Mult.2

\section{Observational data and analysis}

Our data consists of spectra
acquired with UVES at the ESO Kueyen 8.2m telescope,
at a resolution of R$\sim$45\,000.
We have 3 TO stars in the NGC\,6397, 2 TO stars
in NGC\,6752
and the field TO star HD\,218502.
For each cluster star the total integration time
is of the order of 10 hours for each star.
The data has already been described in
\citet{Pasquini04} and \citet{Pasquini07}.
The reduced spectra were downloaded from the ESO archive, 
thanks to the improved strategies for optimal
extraction \citep{ballester}, the S/N ratios
are greatly improved with respect to 
what was previously available.
We measured the EWs of the  \ion{Cu}{i} lines
by fitting a gaussian with
the IRAF task {\tt splot}.
For each star we computed 
a 1D LTE model atmosphere with the
atmospheric parameters given in
\citet{Pasquini04} and \citet{Pasquini07}
and summarised in Table\,\ref{abun}.
We used the ATLAS 9 code \citep{K93a,K05}
in its Linux version \citep{SB04,SB05}.
From this we computed for each line a
curve of growth with the SYNTHE code 
\citep{K93b,K05,SB04,SB05} taking
into account the hyperfine structure
of the lines.
The abundance was determined by interpolation in
these curves of growth. 
For the hydrodynamical models we 
used the \cobold\ code
\citep{Freytag2002AN....323..213F,Freytag2003CO5BOLD-Manual,Wedemeyer2004A&A...414.1121W}
and used spectrum synthesis on these models to 
compute curves of growth and ``3D corrections'',
as defined by \citet{zolfito}, with respect to the 1D LHD models.
The appropriate 3D correction was found for each 
star by interpolating
in the 3D grid. 

\section{Results and conclusions}

From Table \ref{abun} it is immediately clear
that the 3D corrections are large.
The reasons for this can be understood
by looking at Fig.\,1 where the temperature
distribution of one of our 3D models is depicted, 
together with the mean temperature distribution
and that of a corresponding LHD model.
The overcooling is obvious from the top panel
and the contribution functions in the two bottom
panels reflect  the fact that in these cool outer
layers the Cu atoms populate mainly the ground layer,
thus contributing for the bulk of
the absorption of the line, at variance with what
happens in the 1D model (dashed line).
The two bottom panels correspond to two 
different Cu abundances, illustrating how the
tendency to prefer the outer layers
increases with the increasing number of absorbers.
As expected this results in larger 3D corrections
for more metal-rich stars.
However, one should be always cautious when facing
contribution functions like the ones shown in Fig.\,1.
In fact all the computations have been performed in LTE,
it is likely that photons coming from the warm streams
may produce overionisation in these low density outer
layers. If NLTE effects are important a considerable 
resizing of the outer peak of the contribution function
can be expected.
A way to check indications of possible NLTE effects
is to compare the abundances in the cluster, derived from
the resonance lines in TO, with those of giants,
derived from the lines of Mult.\,2.
We derived  abundance in 
NGC\,6397 using the EWs for two giants measured by \citet{gratton82},
for NGC\,6752 we used a UVES spectrum of a giant
star (star  Cl* NGC 6752 YGN 30), already studied by \citet{Yong}.
Neither in 1D nor in 3D giants and dwarfs
provide the same abundance.
In NGC 6752 the dwarf stars imply a higher abundance,
both in 1D and 3D, the reverse is true in NGC 6397.
That the problem is with the modelling of the Cu lines
is confirmed by the analysis of the giant
in NGC\,6752,
for which we were able to measure both
the \ion{Cu}{i} resonance line at 327.3\, nm
and the 510.5\,nm line. Both in 1D and 3D the 
two lines provide abundances which differ by about 0.5\,dex.
We conclude that the  \ion{Cu}{i} resonance lines are not
good abundance indicators, a full 3D-NLTE study should be undertaken
for these lines. At the same time one may  suspect that also
the lines of Mult.\,2 may be affected by deviations from LTE.
The Galactic evolution of copper must be placed on solid grounds
with a better modelling of the line formation.

\balance
 
\begin{acknowledgements}
  We are grateful to L. Pasquini for many useful
  comments on this work.
  We acknowledge financial
  support from EU contract MEXT-CT-2004-014265 (CIFIST).
  We also acknowledge use of the supercomputing centre CINECA,
  which has granted us time to compute part of the hydrodynamical
  models used in this investigation, through the INAF-CINECA
  agreement 2006,2007.
\end{acknowledgements}

\bibliographystyle{aa}

\end{document}